\documentclass[a4paper,12pt]{jpconf}
\usepackage{graphicx}
\usepackage{amsfonts}
\usepackage{cite}
\usepackage{amssymb}
\usepackage{iopams}
\usepackage{amsmath}
\usepackage{xspace} 
\usepackage[%
breaklinks=true,
colorlinks=true,
linkcolor=blue,
urlcolor=blue,
citecolor=blue
]{hyperref}

\begin{document}

\title{Universal scaling relation for growth phenomena}

\author{Evandro A. Rodrigues}
\ead{evandrofisico@gmail.com}
\address{ Instituto de F\'isica, Universidade de Bras\'ilia, 70910-900, Bras\'ilia, DF, Brasil  }

\author{Edwin E. Mozo Luis}
\ead{emozo@id.uff.br}
 \address{Instituto de F\'{\i}sica, Universidade Federal Fluminense,
  Avenida Litor\^{a}nea s/n, 24210-340, Niter\'{o}i, RJ, Brazil}

\author{Thiago A. de Assis}
\ead{thiagoaa@ufba.br}
\address{Instituto de F\'{\i}sica, Universidade Federal da Bahia,
	Campus Universit\'{a}rio da Federa\c c\~ao,
	Rua Bar\~{a}o de Jeremoabo s/n, 40170-115, Salvador, BA, Brazil}

 \address{Instituto de F\'{\i}sica, Universidade Federal Fluminense,
  Avenida Litor\^{a}nea s/n, 24210-340, Niter\'{o}i, RJ, Brazil}

 \author{Fernando A. Oliveira}
\ead{fao@fis.unb.br}
\address{ Instituto de F\'isica, Universidade de Bras\'ilia, 70910-900, Bras\'ilia, DF, Brasil  }
\address{Instituto de F\'{\i}sica, Universidade Federal Fluminense,
  Avenida Litor\^{a}nea s/n, 24210-340, Niter\'{o}i, RJ, Brazil}

\begin{abstract}
The Family-Vicsek relation is a seminal universal relation obtained for the global roughness at the interface of two media in the growth process. In this work, we revisit the scaling analysis and, through both analytical and computational means, show that the Family-Vicsek relation can be generalized to a new scaling independent of the size, substrate dimension $d$, and scaling exponents. We use properties of lattice growth models in the Kardar-Parisi-Zhang and Villain-Lai-Das Sarma universality classes for $1 \leq d \leq 3$ to support our claims. 

\end{abstract}

\section{Introduction} 

The growth phenomenon is a class of dynamic processes happening in the interface between two media that can be described as natural and universal, which encompasses a wide field of applications \cite{MEAKIN93,Barabasi95,krug1997origins,Nath18}.  
Some examples of growth systems are atomic deposition~\cite{Csahok92}, evolution of bacterial colonies~\cite{Ben-Jacob94, Matsushita90}, spherical models~\cite{Henkel15},  and corrosion~\cite{Mello01, Renner06,Reis03,Rodrigues15,Mello15,Alves16,Gomes19}. Models have been proposed and studied through very creative  experiments~\cite{Ben-Jacob94, Matsushita90,Miranda10}, analytical and numerical solutions \cite{Wio10a,Wio10b}, simulations~\cite{Reis04,Reis05,Buceta14,JAP2023}, and few exact results for specific situations \cite{edwards1982surface,Kardar86,Dotsenko10,Calabrese10,Amir11,Sasamoto10,Doussal16}. 

In the context of lattice models, different internal dynamics can lead to distinct irregular interfaces characterized by measures like the mean and the standard deviation of the surface height. The latter is often called the global roughness, $w(t)$ whose square is defined as 

\begin{equation} w^2(L,t) = \frac{1}{L^{d}} \sum_{i=1}^{L^d} {\chi_i}^2(t),
\label{w0}
\end{equation}
where $\chi_i(t) \equiv {h_i}(t) - <h_i(t)>$. Here, $h_i(t)$ is the height of the $i$-th site, with $i=1,2...L^{d}$, where $d$ is the substrate dimension.
Though the average height $<h_i(t)>$ increases continuously due to the growth process, the dynamic in some interface growth models leads to surface width saturation after the roughening buildup. For many systems, the saturated surface width, $w_s$, is often expressed as a function of the lateral size of the substrate, $L$, as the power law $w_s\sim L^{\alpha}$, $\alpha$ being the global roughness exponent. The saturation occurs at times larger than a characteristic time $t_\times$ that follows the power law $t_\times \sim L^{z}$, where $z$ is the dynamic exponent. Before saturation ($t\ll t_\times$), $w(L,t)$ evolves as a power law $w(L,t) \sim t^{\beta}$, where $\beta$ is the growth exponent.

In the context of continuum models, most surface growth phenomena are modeled through stochastic Langevin equations, such as the Kardar-Parisi-Zhang (KPZ) one \cite{Kardar86} 

\begin{equation}
\label{KPZ}
\dfrac{\partial h(\vec{x},t)}{\partial t}=\mu \nabla^2 h(\vec{x},t) +\dfrac{\lambda}{2}[\vec{\nabla}h(\vec{x},t)]^2 + \xi(\vec{x},t).
\end{equation}
Here $h: \mathbb{R}^d \times \mathbb{R}^+ \rightarrow \mathbb{R}$ {denotes} the interface height at point $\vec{x} \in \mathbb{R}^d$ at time $t\geqslant0$. The parameters $\mu$ (surface tension) and $\lambda$ are related to Laplacian smoothing and tilting, respectively. The growth appears in a direction orthogonal to the $d$-dimensional substrate hyperplane defined by $\vec{x}$, so we are dealing with a $(d+1)$-dimensional space. For applications in film growth, $d=2$. We note that $h(\vec{x},t)$ displays different scaling properties in the growth direction from those along directions in the $\vec{x}$ plane. The stochastic process is characterized by the zero mean white noise, $\xi(\vec{x},t)$,  with 

\begin{equation}
    \left\langle \xi(\vec{x},t) \xi(\vec{x}',t')\right\rangle = 2D\delta^{(d)}(\vec{x}-\vec{x}')\delta(t-t'),
\end{equation} 
where $D$ is the noise intensity. %\color{red} The last relation is sometimes called the Fluctuation-Dissipation theorem ~\cite{Barabasi95,edwards1982surface,GomesFilho21,Anjos21} \color{black}.

The hypothesis that we can treat a macroscopic system as being homogeneous brings us to the scaling-invariance and universality of the scaling exponents~\cite{Kadanoff00}. Off-equilibrium lattice dynamics also exhibit important scaling phenomena~\cite{Oliveira95,Oliveira00}. These concepts are particularly important in phase transition~\cite{Kadanoff00} and growth phenomena~\cite{Kardar86}. For surface growth, we shall discuss the Family-Vicsek relation and its generalization, our foremost goal in this work.

For most growth models, there is no exact analytical result for  $w(t)${, the EW case being a rare exception due to its linearity}. Indeed, if we consider the particular
case of the KPZ class, one of the most studied growth models. It has not been exhausted as yet. For $d=1$, probabilities distributions  has been obtained \cite{Dotsenko10,Amir11,Calabrese10,Doussal16} for the height deviations $y=h-<h>$ in the form $P(y/w)$, which does not allows one to obtain $w(t)$. To obtain $w(t)$, we need to go far beyond our present knowledge. Consequently, we need the support of computer simulations for a while.

Before proceeding with the simulations, we note that the growth process creates a fractal interface~\cite{Barabasi95, krug1997origins,Luis22,Luis23,GomesFilho21b}. Second,  the self-affinity relationship within fractals is equivalent to the scale dynamics in the renormalization process. Finally, scaling yielded the first nontrivial relations between exponents. Thus  we  write \cite{Kadanoff00}

\begin{equation}
\label{WL}
w(L,t)=\rho^{-1} w(\rho^b L,\rho^c t),
\end{equation}
where $\rho$ is an arbitrary parameter. To get a useful relation we choose $\rho=L^{-1/b}$, which imply that $b \equiv \alpha^{-1}$, and $c \equiv \beta^{-1}$. Therefore, we obtain

\begin{align}
\label{Sc1}
w(L,t)&= w_s  w(1,\tau) \nonumber\\
 &= w_s f_1(\tau)=
  \begin{cases}
 c_1t^\beta , &\text{ if~~ } t \ll t_\times\\
 w_s, &\text{ if~~ } t \gg t_\times,\\
 \end{cases}
\end{align}
known as Family-Vicsek (FV) relation \cite{Family85}. Here, $c_1$ is a constant and $\tau \equiv t/t_x$. The FV relation reflects the results found in many experimental situations \cite{Barabasi95,SR2011,PRL2020}. 

A consequence of this scaling is that for a given substrate dimension $d$, the curves $w(L,t)/w_s$, as a function of $\tau$, collapse into the same curve independent of $L$. One of the consequences of the FV relation  is the scaling
\begin{equation}
\label{FS}
z=\frac{\alpha}{\beta},
\end{equation}
which is universal for any growth and dimension, i.e., is model independent~\cite{Barabasi95}.

\section{Generalization of the Family-Vicsek relation} 

Now we argue that there is a hidden, more universal relation that does not depends on the substrate dimension $d$.  
Choosing $\rho=t^{-1/c}$ , then Eq. (\ref{WL}) becomes 
\begin{equation}
\label{WL2}
w(L,t)=t^\beta w( [L^z/t]^{1/z},1)=t^\beta f_2(\tau),
\end{equation} 
with the same exponents $b$ and $c$ as in Eq. (\ref{Sc1}). Here
\begin{equation}
\label{f2}
f_2(\tau)= 
\begin{cases}
c_1, &\text{ if~~ } \tau \ll 1\\
w_s/\tau^\beta, &\text{ if~~ } \tau \gg 1.\\
\end{cases}
\end{equation}
Here $c_1$ is as in Eq. (\ref{Sc1}). Second, let us divide Eq. (\ref{WL2}) by $w_s$,  and considering that $w_s=L^\alpha=t_\times^\beta$, 
    $\Omega \equiv \left[w(L,t)/w_s\right]^{1/\beta}$ and  $\psi(\tau) \equiv \tau f_2(\tau)^{1/\beta}$ to get
\begin{equation}
\label{SC2}
\Omega=\psi(\tau)= 
\begin{cases}
\tau , &\text{ if~~ } \tau \ll 1\\
1, &\text{ if~~ } \tau \gg 1.\\
\end{cases}
\end{equation}
The function $\psi$ is universal for each universality class since it does not depend on $d$, $L$, or the scaling exponents.
Thus Eq. (\ref{SC2})  is a generalization of the FV relation (\ref{Sc1}).
In brief, the relation (\ref{Sc1}) gives us scaling functions, which are independent of the size $L$, while the relation  (\ref{SC2}) gives us functions that are size, dimension, and exponents independent. { Note that in the FV relations we scale $w(t)/L^{\alpha}$ and $t/L^z$. In the generalized FV relation, we add a new scale for the exponent $\beta$, i.e., $[w(t)/L^{\alpha}]^{1/\beta}$.  Thus, we get the missing relation and contemplate all the exponents in this way.} This is our major result.

\begin{figure}
	\centering
	\includegraphics[scale=0.9]{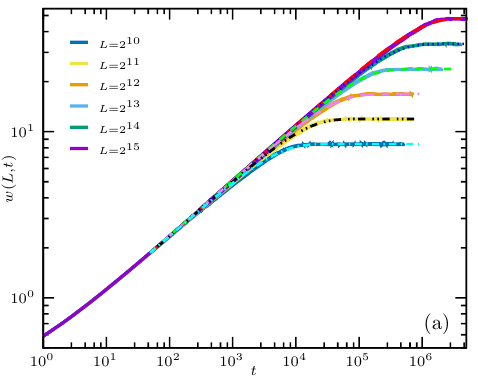}\\
    \includegraphics[scale=0.9]{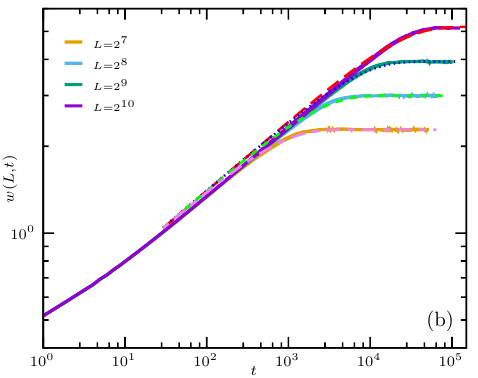}
	\caption{$w(L,t)$ vs. $t$ considering the RSOS model and $L=2^{n}$. (a) Results for $d=1$, with  $10 \leq n \leq 15$, considering $12000$, $10000$, $8000$, $4800$, $3200$ and $3000$ independent realizations, respectively.  (b) Results for $d=2$ with  $7 \leq n \leq 10$, considering $9600$, $7200$, $4800$ and $3000$ independent realizations, respectively. The dashed lines are fittings using Eq. (\ref{wff}).}
	\label{figRSOS}
\end{figure}

\begin{figure}
	\centering
	\includegraphics[scale=0.9]{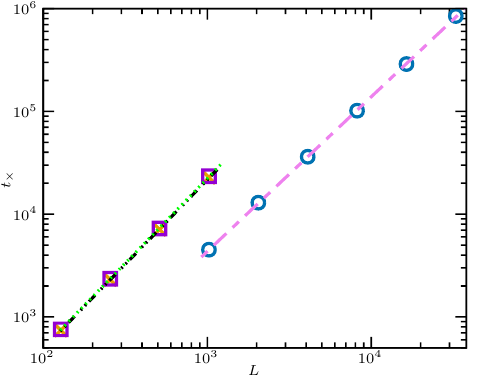}
	\caption{$t_{\times}$ vs. $L$ for RSOS model in $d=1$ (circles) and $2$ (squares and crosses), where $t_{\times}$ is extracted from Eq. (\ref{wff}). The dashed lines have a slope $1.505(5)$ [if $\beta = 1/3$ is considered] (violet), and $1.647(9)$ [if $\beta = 0.2361$ is considered ~\cite{GomesFilho21b}] (green) and $1.643(8)$ [if $\beta = 0.2414$ is considered ~\cite{Kelling17}] (black).}
	\label{txRSOS}
\end{figure}

Observe that,  as in the FV case, we do not need to have $w(t)$. Thus we do not have $\psi(\tau)=\Omega(\tau)=\left[ w(L,t)/w_s \right]^{1/\beta}$ to finalize our scaling relation.  However, it would be useful to have at least an approximation for $\psi(\tau)$. 
We shall comment here that the analysis from Alves \textit{et al.} \cite{Alves16},  for etching model in $d=1$ yields
	\begin{equation}
	\label{wf}
	w(L,t)= w_s\left[1-\exp{(-\tau})\right]^{1/2},
	\end{equation}
 for times $t <t_c< t_\times$, where $t_c$  is the time where correlations start, i.e., before that, we have a simple Brownian motion. Additionally, they obtained $w_s$ and showed that $w_s \propto L^{1/2}$, i.e., $\alpha=1/2$ exactly, which supports their method and the conclusion that the etching model belongs to the KPZ universality class. As we already mentioned, that conclusion was later proved by Gomes \textit{et al.}~\cite{Gomes19} for all dimensions. 
Based on that, we propose the empirical relation
 \begin{equation}
	\label{wff}
	w(L,t)= w_s\left[1-\exp{(-\tau})\right]^{\beta}.
	\end{equation}
It means that we replace the uncorrelated exponent of Eq. (\ref{wff}) with the correlated one, which satisfies the scaling (\ref{Sc1}) and (\ref{SC2}). In this way, we get
\begin{equation}
\label{psi}
\psi(\tau)=1-\exp(-\tau),
\end{equation}
which shows up to be precise in treating the empirical data.

\section{Simulations of KPZ lattice models}

Next, to verify if Eq. (\ref{psi}) is reasonable, we simulate three cellular automata that consistently model properties of the KPZ equation \cite{Kardar86}, namely: the restricted solid-on-solid (RSOS) \cite{kim89}, the single step (SS) \cite{PRA86, PRB87, PRB88, PRE2000, Daryaei20} and the etching models \cite{Mello01}. Recently, Gomes \textit{et al.} \cite{Gomes19} have shown that the $d+1$ dimensional etching model yields the KPZ equation in the continuous limit. 
For the particular case of the KPZ models, the fluctuation-dissipation relation has recently obtained quite considerable attention \cite{Anjos21,GomesFilho21}.

\begin{figure}
\centering
	\includegraphics[scale=0.85]{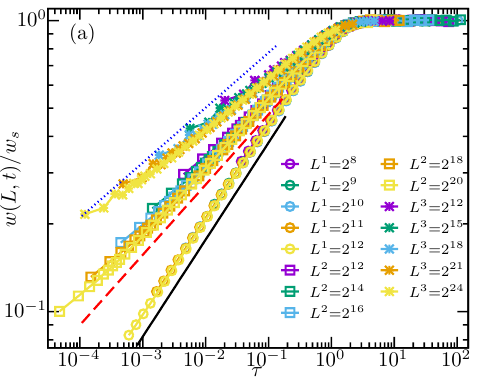}
	\includegraphics[scale=0.85]{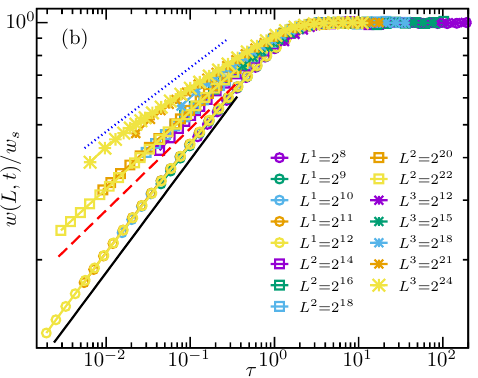}
 	\includegraphics[scale=0.85]{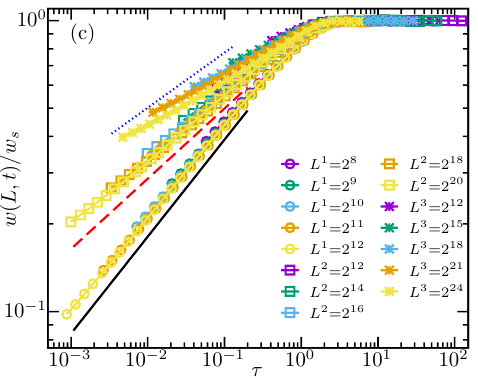}
   
	\caption{$w(L,t)/w_s$ vs. $\tau$, considering the (a) RSOS, (b) etching, and (c) SS ($p=1$) models, for $L^{d} = 2^{nd}$ with $1 \leq d \leq 3$. For $d=1$, the RSOS, etching, and SS models were simulated considering $ 8 \leq n \leq 12$ with $16000$, $14000$, $12000$, $10000$ and $8000$ independent realizations, respectively. For $d=2$ the RSOS and SS models were simulated for $ 6 \leq n \leq 10$, with $4000$, $1600$, $1200$, $900$ and $700$ independent realizations, respectively. The etching model was simulated for $7 \leq n \leq 11$  with $1600$, $1200$, $1000$, $700$ and $500$ independent realizations, respectively. For $d=3$, the RSOS model was simulated for $4 \leq n \leq 8$ with $10000$, $8000$, $6000$, $4000$ and $3200$ independent realizations, respectively.  etching model was simulated for $4 \leq n \leq 8$ with $10000$, $8000$, $6000$, $4000$ and $3200$ independent realizations, respectively. The SS $(p=1)$ model was simulated for $4 \leq n \leq 8$ with $10000$, $8000$, $6000$, $2520$ and $120$ independent realizations, respectively. Filled (black), dashed (red), and dotted (purple) lines indicate slopes consistent with the KPZ growth exponents in $d=1$, $2$, and $3$, respectively, reported in Ref. \cite{Oliveira2022}.}
	\label{figcoldim}
\end{figure}

\begin{figure}
\centering
	\includegraphics[scale=0.85]{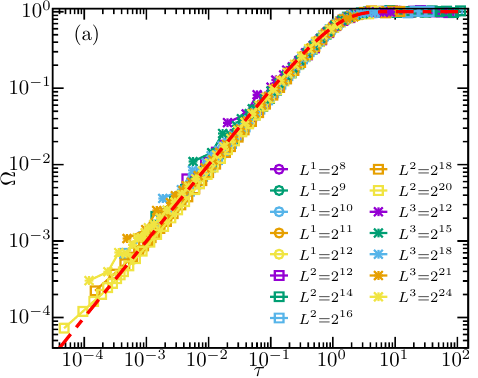}
	\includegraphics[scale=0.85]{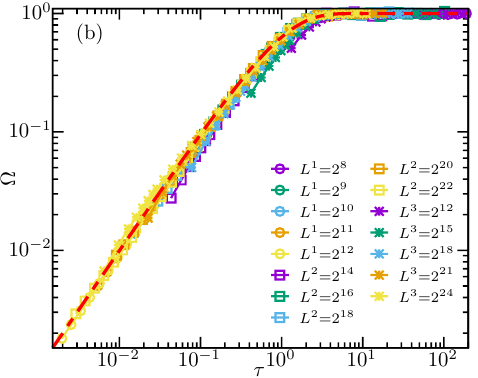}
 	\includegraphics[scale=0.85]{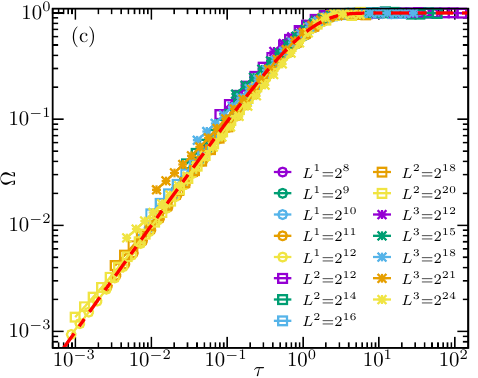}
   
	\caption{ $\Omega$ vs. $\tau$, where $\Omega = \left[w(L,t)/w_{s} \right]^{1/\beta}$ and $\tau = t/t_{\times}$. The results are presented for $L$ and the number of independent realizations as the data presented in  Fig. \ref{figcoldim}, considering the (a) RSOS, (b) etching, and (c) SS ($p=1$) models. The dashed line red is the fit of Eq. (\ref{psi}). The plot exhibit independence in $L$ and $d$}
	\label{figOmega0}
\end{figure}

In the simulations, we randomly choose, at time $t$, a column of the deposit in the $R^{d}$ space. We consider the lattice constant the unit length (i.e., $a = 1$) and the flux of particles per site per
time unit towards the substrate be constant, so a (dimensionless) time unity corresponds to the deposition of $L^{d}$ particles. All simulations have been performed by considering periodic boundary conditions and an initial substrate with $h(\vec{x},0)=0$, except with the SS model, for which $h_i= [1+(-1)^i]/2 $ for $d=1$, $h_{i,j}= [1+(-1)^{i+j}]/2 $ for $d=2$ [$ 1 \leqslant (i,j)  \leqslant L$] and $h_{i,j,k}=  [1+(-1)^{i+j+k}]/2 $ for $d=3$ [$ 1 \leqslant (i,j,k)  \leqslant L$], with $\left\{i,j,k\right\} \in \mathbb{N}$.

\begin{figure}
	\centering
       \includegraphics[scale=0.85]{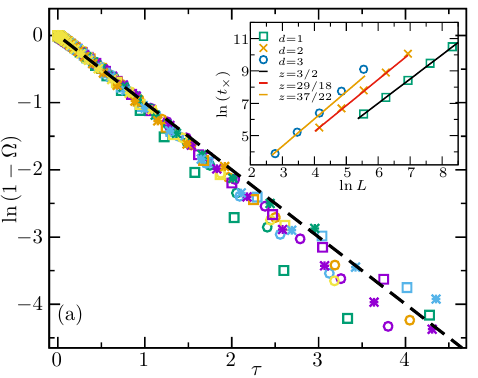}
        \includegraphics[scale=0.85]{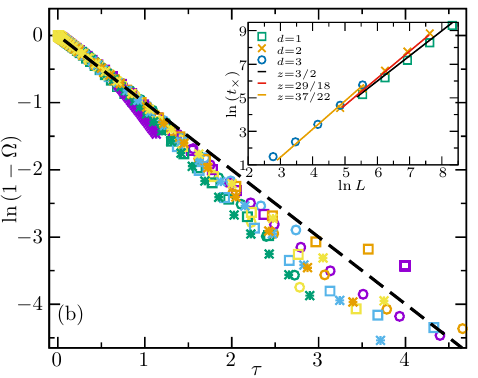}
         \includegraphics[scale=0.85]{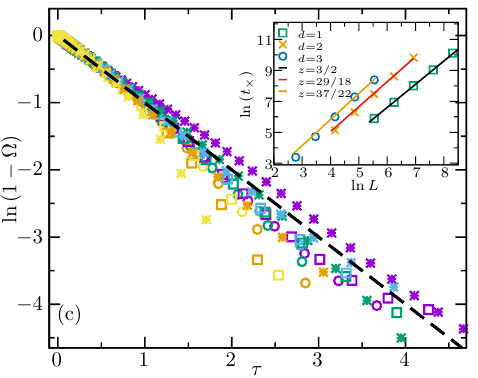}

	\caption{{$\ln{(1- \Omega)}$ as function of  $\tau$, where $\Omega = \left[w(L,t)/w_{s} \right]^{1/\beta}$ and $\tau = t/t_{\times}$. The results are presented for $L$ and the number of independent realizations as the data presented in  Fig. \ref{figcoldim}. We consider (a) RSOS, (b) etching, and (c) SS $(p=1)$ models. The dashed black line corresponds to $f(\tau) = -\tau$. The insets show the evolution of $\ln{(t_\times)}$ as function of $\ln{L}$, where $t_\times$ is extracted of the Eq. (\ref{wff}). The values of $\beta$ and $z$ are the KPZ exponents for $1\leq d \leq 3$ reported in \cite{Oliveira2022}.}}
	\label{figOmega}
\end{figure}

In the deposition version of the etching model~\cite{Mello01,Gomes19}, as considered here, after each deposition attempt, the height of the incidence column, $h_0$, was increased by one lattice unit, and any neighboring column smaller than $h_0$ increased until its height was $h_0$. In the RSOS model, an incident particle may stick at the top of the incidence column if the local slope of the incidence column and each of the neighboring columns does not exceed one lattice unit. Otherwise, deposition is rejected. In the SS model, at any time $t$, if the column height of a randomly selected site is a local minimum, then it is increased by two lattice parameters with probability $p$. Otherwise, if we have a maximum, the column height is decreased by two lattice parameters with probability $1-p$. In this work, we present the results for the SS model considering $p=1$.

Before continuing, we will discuss the proposed Eq. (\ref{wff}). In Fig. \ref{figRSOS}, we plot $w(t)$ as a function of $t$ for the  RSOS model. In (a), the results are shown for $d=1$, where the exact KPZ growth exponent is $\beta=1/3$. Large lateral sizes, $L$, and a reasonable number of independent realizations are considered. In (b) for $d=2$, we use the  result of Kelling \textit{et al.}~\cite{Kelling17} $\beta=0.2414(2)$. The data produced for both dimensions are precise. Thus, we adjust $w_s$ and $t_{\times}$ to Eq. (\ref{wff}). The dashed line is the curve obtained from Eq. (\ref{wff}). It superimposes the points obtained by the simulation. 

In the Fig. \ref{txRSOS} we plot $t_{\times}$ as function of $L$ for the  RSOS model.  We use the $t_{\times}$ from  the data of Fig. \ref{figRSOS}.  The figure exhibit the law $t_{\times} \propto L^z$, for $d=1,2$.  From the  slope we obtain $z=1.505(5)$ for $d=1$, in good agreement with the exact value $z=3/2$. For $d=2$ we obtain $z=1.643(8)$, with a lower precision than in the one-dimensional case, but compatible with the best simulations~\cite{Kelling17} $z=1.6111(3)$.
 In this way, both Fig. \ref{figRSOS} and Fig. \ref{txRSOS} support the guess function Eq. (\ref{wff}).
Obtaining $t_{\times}$, and consequently $z$, is a difficult task in computational physics. In this way, both $z$ and collapse are important achievements.   Note that Eq. (\ref{wff}) is not necessary for the scaling. On the other hand, if it works, it is good support for data analysis and  scaling.

In Fig. \ref{figcoldim}, we show the results for $w(L,t)$, in units of $w_s$, as a function of $t$, in units of $t_\times$, for the models described before for $d + 1$ dimensions,
 with $d =1,2$ and $3$.  
It can be seen that changing the scale from $w(L,t) \rightarrow w(L,t)/w_s$ and $ t \rightarrow t/t_\times$ already collapses the curves of each dimension without further modification. For each dimension, $d$, data collapses, i.e., curves with different sizes $L = 2^n$ are re-scaled to a single curve. This is the soul of the FV relations. We show the results for the RSOS [Fig. \ref{figcoldim} (a)], etching [Fig. \ref{figcoldim} (b)] and SS ($p=1$) [Fig. \ref{figcoldim} (c)] models, respectively. Of course, since small values of $L$ have been considered, finite-size effects are expected.
Oliveira \cite{Oliveira2022} proposed that, for a given substrate dimension $d$, the KPZ growth exponent is given by $\beta^{\mathrm{KPZ}}(d) = 7/\left(8d+13\right)$, which agrees well with our results shown in Fig. \ref{figcoldim} for the three lattice models considered here. 
 
 In Fig. \ref{figOmega0}, we show the results for dimensionless $\Omega(\tau)=[w(L,t)/w_s]^{(1/\beta)}$, as a function of $\tau= t/t_\times$, for the models described in Fig. \ref{figcoldim}, for dimensions $d + 1$, with $d =1,2$ and $3$.  Note that the collaps is independent of dimension and size. Of course, the finite-size effects that appear in Fig. \ref{figcoldim} are expected.

Figures \ref{figOmega} (a), (b) and (c) show $\ln{(1- \Omega)}$, as a function of $\tau$ for RSOS, etching, and SS models, considering $1 \leq d \leq 3$. The data analysis suggests that Eq. (\ref{psi}) fits well with the numerical results. However, at saturation where $w \approx w_s(1 \pm \epsilon)$, with $\epsilon$ as a relative error, $\Omega \approx 1 \pm \epsilon/\beta$, i.e., the error is amplified by a factor $1/\beta \geq 3$. {Furthermore, the $\ln(1-\Omega)$ becomes $\ln(\epsilon/\beta)$, which are large for small $\epsilon$; for example for $\tau=4$, we have $\epsilon=0.0183 \beta$ . Furthermore, the fluctuation becomes large for $w \approx w_s $, and what we see is just fluctuations which appears for the large  values of $\tau$.} Albeit all curves are collapsed and show the same general behavior, each model has, however, some specific behavior at the beginning of the growth phase. Interestingly, this good collapse is observed if we consider the KPZ scale exponents recently reported in Ref. \cite{Oliveira2022}. Note that the curve $\psi(\tau)$ does not depend on size and dimension, so it must be universal in this regard. This result supports (\ref{SC2}) as a generalization of the FV relation. Equation (\ref{psi}) agrees with the collapse of data such as emerges in Fig. \ref{figOmega} if one forgets the slight deviation of each model, as we point out before. Finally, we can note the remarkable agreement between the dynamical exponent of KPZ reported in \cite{Oliveira2022} [i.e. $z^{\mathrm{KPZ}}(d)$ =$\left(8d+13\right)/\left(4d+10 \right)$] and the one extracted from $\ln{(t_\times)}$ as a function of $\ln{(L)}$, where $t_\times$ is extracted from the equation. (\ref{wff}) (see the insets of Fig. \ref{figOmega}).

\section{Villain-Lai-Das Sarma (VLDS) lattice model}

We extended our analysis to the VLDS class, where diffusion is the dominant mechanism during growth. In that case, $h(\vec{x},t)$ is the solution of the VLDS equation ~\cite{Barabasi,Villain91,Lai91}:
\begin{equation}
\label{VLDS}
\dfrac{\partial h(\vec{x},t)}{\partial t}=-\nu_4 \nabla^4 h(\vec{x},t) +\lambda_4 \nabla^2[\vec{\nabla}h(\vec{x},t)]^2 + \xi(\vec{x},t),
\end{equation}  
where $\nu_{4}$ and $\lambda_{4}$ are constants, and $\xi(\vec{x},t)$ is a nonconservative Gaussian noise, as previously described. The incoming particle flux is omitted in Eq. (\ref{VLDS}). From renormalization group (RG) analysis, the scaling exponents $\beta$, $\alpha$, and $z$ for the VLDS equation are given by: 
\begin{equation}
\label{alf0}
\alpha^{\mathrm{VLDS}}(d) = \frac{4-d}{3}-\zeta(d),
\end{equation} 
 and $z^{\mathrm{VLDS}}(d) = (d + 8)/3-2\zeta(d)$. Of course, $\beta^{\mathrm{VLDS}}(d) = \alpha^{\mathrm{VLDS}}(d)/z^{\mathrm{VLDS}}(d)$. For a two loops RG analysis \cite{Janssen97},
\begin{equation}
\label{deltavlds}
\zeta(d)=0.01361\left(2-\frac{d}{2}\right)^2.
\end{equation}

\begin{figure}
\centering
	\includegraphics[scale=0.85]{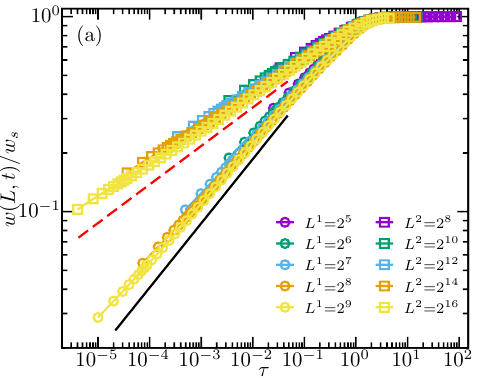}
	\includegraphics[scale=0.85]{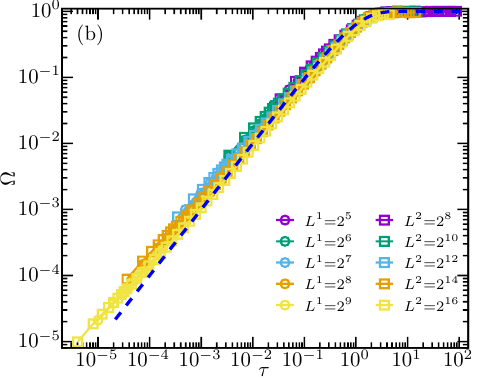}\\
    \includegraphics[scale=0.85]{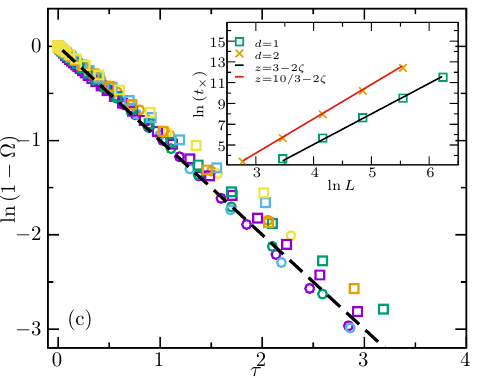}
	
	\caption{ (a) $w(L,t)/w_s$ vs. $\tau$, considering the  CRSOS model, for $L^{d} = 2^{nd}$ with $1 \leq d \leq 2$. For $d=1$, were simulated considering $ 5 \leq n \leq 9$ with $22000$, $20000$, $18000$, $16000$ and $14000$ independent realizations, respectively. For $d=2$, were simulated considering $ 4 \leq n \leq 8$ with $10000$, $6000$, $4000$, $1600$ and $1200$ independent realizations, respectively. Filled (black) and dashed (red) lines indicate slopes consistent with the VLDS growth exponents considering two-loop correction in $d=1$ and $2$, respectively, reported in Ref. \cite{Janssen97}. (b) $\Omega$ vs. $\tau$, with $\Omega = \left[w(L,t)/w_{s} \right]^{1/\beta}$ and $\tau = t/t_{\times}$, where  the dashed blue line is the fit of Eq. (\ref{psi}).  (c) $\ln{(1- \Omega)}$ as a function of  $\tau$. The results are presented for $L$ and the number of independent realizations as the data presented in  (a). The dashed black line corresponds to $f(\tau)=-\tau$. The inset show the evolution of $\ln{(t_\times)}$ as function of $\ln{L}$, where $t_\times$ is extracted of the Eq. (\ref{wff}). The values of $\beta$ and $z$ are the VLDS exponents for $d=1$ and $2$ with two-loop correction, where $\zeta = 0.0306$ ($d=1$) and $0.0136$ ($d=2$), as reported in Ref. \cite{Janssen97}. } 
	\label{figCRSOS}
\end{figure}
In Fig. \ref{figCRSOS}, we show the results for $w(L,t)/w_s$, as a function of $\tau$, for the conservative restricted solid on solid (CRSOS) model ~\cite{Kim94, Kim97} for $d =1,2$. This model is connected with the VLDS universality class. In the CRSOS model, one site is randomly selected for one adatom deposition. The height differences $\delta h$ between nearest-neighbors obey the restriction $\delta h \leqslant \delta H_{max}$. This work presents the results for the case $\delta H_{max}=1$. If this condition is satisfied for the randomly chosen incidence site, the particle remains permanently stuck there. 
Otherwise, it searches for the nearest position where the condition is satisfied, which then becomes the location of the deposition. In the case of multiple options, one of them is randomly chosen. Again, the transformation $w(L,t) \rightarrow w(L,t)/w_s$ and $t \rightarrow \tau$ collapses curves of each dimension, for $d=1,2$, as shown in Fig. \ref{figCRSOS} (a).

Figure \ref{figCRSOS} (b) shows {$\ln{(1- \Omega)}$}, as a function of $\tau$ for the CRSOS model for $d=1,2$. The data analysis suggests that   Eq. (\ref{psi}) also fits the numerical results for the VLDS model. Interestingly, this good collapse is observed if we consider the VLDS scale exponents provided from a two-loop RG analysis \cite{Janssen97}. This result supports (\ref{SC2}) as a generalization of the FV relation extending our findings for VLDS class. Finally, one can note the remarkable agreement between the VLDS dynamical exponent reported in \cite{Janssen97} and the one extracted from $\ln{(t_\times)}$ as a function of $\ln{L}$, where $t_\times$ is extracted of the Eq. (\ref{wff}) [see the insets of Fig. \ref{figCRSOS} (b)].

\section{Conclusions} 

In conclusion, we have shown that the Family-Vicsek (FV) relation  (\ref{Sc1}) can be generalized to a universal relation, i.e., the generalized FV relation (\ref{SC2}), which is size and dimension independent, regardless of the actual form of the function $\psi( \tau)$ being unknown.
We support our argument with simulations of the
KPZ and VLDS models, where the collapse of data for different sizes and dimensions steam naturally from that. Our results suggest as well the most straightforward form for the function $\psi(\tau)$ of relation (\ref{SC2})  as exhibited in Eq. (\ref{psi}). We use analytical arguments to show that Eq. (\ref{psi}), nevertheless not exact, is an excellent function to obtain the collapse of the data and the dynamic exponents for the KPZ and VLDS classes. Remarkably, our results show that the normal scaling is expected for VLDS class \cite{Reis2013,Edwin1,Edwin2}, contrasting with some claims of anomalous scaling for such class \cite{XIA2013}.

\section{Acknowledgements}
This work was supported by the Conselho Nacional de Desenvolvimento Cient\'{i}fico e Tecnol\'{o}gico (CNPq), Grant Nos. 312497/2018-0 (FAO) and 310311/2020-9 (TAdA), and the Funda\c{c}\~ao de Apoio a Pesquisa do Distrito Federal (FAPDF), Grant No. FAPDF- 00193-00000120/2019-79. TAdA and FAO also are grateful for funding from the Funda\c{c}\~{a}o de Amparo \`{a} Pesquisa do Estado do Rio de Janeiro (FAPERJ), Grants Nos. E-26/203.860/2022 and E-26/203.953/2022, respectively. TAdA is grateful to the National Laboratory for Scientific Computing (LNCC, Brazil). This study was financed in part by the Coordena\c{c}\~{a}o de Aperfei\c{c}oamento de Pessoal de N\'{i}vel Superior - Brasil (CAPES) - Finance Code 001.

\vspace{1.0cm}

\section*{References} 
\vspace{0.5cm}

%\bibliography{Rf_Family_ViN}

\providecommand{\newblock}{}

\end{document}